\documentclass[prd,reprint,showpacs,showkeys]{revtex4-1}
\usepackage{amsfonts}
\usepackage{amssymb}
\usepackage{amsmath}
\usepackage{graphicx}
\usepackage[font={footnotesize,it}]{caption}

\input{tcilatex}

\begin{document}

\title{Black hole solution in third order Lovelock gravity has no
Gauss-Bonnet limit}
\author{Z. Amirabi}
\email{zahra.amirabi@emu.edu.tr}
\affiliation{Department of Physics, Eastern Mediterranean University, G. Magusa, north
Cyprus, Mersin 10 - Turkey}
\date{\today }

\begin{abstract}
We revisit the spherically symmetric third order Lovelock black hole
solution in 7-dimensions. We show that the general solution for the metric
function does not admit the Gauss-Bonnet (GB) limit. This is not expected
due to the linear superposition of the second (GB) and third order Lovelock
Lagrangians in the general action. It is found that the two branches of the
GB solutions are indeed the limit of the other two complex solutions of the
field equations in the third order Lovelock gravity. These two complex
solutions could not be accepted as the solutions of the Einstein's field
equations which are supposed to be real values function on entire real $r-$%
axis. A new solution which is only valid if the third order Lovelock
parameter is small is introduced which can be considered as the natural
extension of the general relativity (GR) to the third order Lovelock
modified theory of gravity. We also generalize the discussion to the higher
dimensional third order Lovelock gravity coupled to the matter sources with
cosmological constant.
\end{abstract}

\maketitle

\section{Introduction}

The modified theory of gravity has been introduced even before the Lovelock
theory \cite{1}. The advantage of the Lovelock theory \cite{2} is to keep
the field equations to second order (derivative of metric function), as the
original idea of Einstein's theory. In addition the Lovelock theory is
ghost-free \cite{3} and a generalized Birkho theorem is also valid there 
\cite{4}. Unlike Einstein's theory, the modified theory of gravity proposed
by Lovelock \cite{2} includes the higher order of curvature and in its most
general form its Lagrangian is given by%
\begin{equation}
\mathcal{L}=\sum_{s=0}^{\left[ \frac{d-1}{2}\right] }\alpha _{s}\mathcal{L}%
_{s}
\end{equation}%
in which $\alpha _{s}$ and $\mathcal{L}_{s}$ are the $s-$order Lovelock
parameter and Lagrangian respectively. For instance, $\alpha _{0}=-\frac{%
\left( d-1\right) \left( d-2\right) }{3}\Lambda $ and $\mathcal{L}_{0}=1,$ $%
\alpha _{1}=1$ and $\mathcal{L}_{1}=R,$ $\alpha _{2}=$Gauss-Bonnet parameter
and 
\begin{equation}
\mathcal{L}_{2}=R_{\mu \nu \gamma \sigma }R^{\mu \nu \gamma \sigma }-4R_{\mu
\nu }R^{\mu \nu }+R^{2},
\end{equation}%
$\alpha _{3}=$third order Lovelock parameter and 
\begin{multline}
\mathcal{L}_{3}=2R^{\mu \nu \gamma \tau }R_{\gamma \tau \xi \lambda
}R_{\;\;\mu \nu }^{\xi \lambda }+8R_{\text{ \ \ \ }\xi \lambda }^{\mu \nu
}R_{\text{ \ \ \ }\nu \kappa }^{\xi \tau }R_{\text{ \ \ \ }\mu \tau
}^{\lambda \kappa } \\
+24R^{\mu \nu \gamma \tau }R_{\gamma \tau \nu \lambda }R_{\text{ \ \ }\mu
}^{\lambda }+3RR^{\mu \nu \gamma \tau }R_{\gamma \tau \mu \nu } \\
+24R^{\mu \nu \gamma \tau }R_{\gamma \mu }R_{\tau \nu }+16R^{\mu \nu }R_{\nu
\tau }R_{\text{ \ \ }\mu }^{\tau }- \\
12RR_{\mu \nu }R^{\mu \nu }+R^{3}
\end{multline}%
and so on. In particular, the second order (which is known as the
Gauss-Bonnet theory of gravity \cite{5}) and the third order of Lovelock
gravity have received intensive attentions during the last decades \cite{6}.

In this Brief Report, our main concern is the black hole solution in static
spherically symmetric third order Lovelock gravity which was first
introduced in \cite{7} and ever since has been studied in different aspects 
\cite{8}. We shall show in this study that the only real solution for the
Einstein equations in third order Lovelock gravity fails to give the second
order Lovelock black hole solution (GB solution) when the third order
Lovelock parameter $\alpha _{3}$ is set to zero (This has been indirectly
noted by Wheeler in \cite{9} and later on by Myers and Simon in \cite{10}).
This indeed does not mean that the solution found is not correct but perhaps
this signals that this solution is not a solution one may wish to find. We
also show that the correct GB limit of the solution comes from the complex
solutions of the third order Lovelock gravity. Finally we introduce a new
solution to the third order Lovelock gravity which is acceptable for the
small values of $\alpha _{3}$ and it has the correct GB limit. Our analysis
is first given in $7-$dimensions vacuum without cosmological constant and a
trivial generalization to $d-$dimensional with cosmological constant and
matter source will be determined after.

\section{$7-$dimensional third order Lovelock gravity}

We start with the action of the third order Lovelock gravity in $7-$%
dimensions ($16\pi G=1$)%
\begin{equation}
I=\int d^{7}x\sqrt{-g}\left( R+\alpha _{2}\mathcal{L}_{2}+\alpha _{3}%
\mathcal{L}_{3.}\right) ,
\end{equation}%
in which the Ricci scalar is shown with $R$ which is also the first order
Lovelock Lagrangian, $\mathcal{L}_{2}$ is the second order Lovelock
Lagrangian which is the well known Gauss-Bonnet term and finally the third
order Lovelock Lagrangian is shown by $\mathcal{L}_{3.}$ Also, $\alpha _{2}$
and $\alpha _{3}$ are the Gauss-Bonnet and third order Lovelock parameters.
The static spherically symmetric line element is given by%
\begin{equation}
ds^{2}=-f\left( r\right) dt^{2}+\frac{dr^{2}}{f\left( r\right) }%
+r^{2}d\Omega _{5}^{2}
\end{equation}%
where $f\left( r\right) $ is given by the Einstein equations 
\begin{equation}
\mathcal{G}_{\mu \nu }^{\left( 1\right) }+\alpha _{2}\mathcal{G}_{\mu \nu
}^{(2)}+\alpha _{3}\mathcal{G}_{\mu \nu }^{\left( 3\right) }=0
\end{equation}%
in which $\mathcal{G}_{\mu \nu }^{\left( 1\right) }=G_{\mu \nu }$ is the
Einstein tensor and $\mathcal{G}_{\mu \nu }^{(2)}$ and $\mathcal{G}_{\mu \nu
}^{\left( 3\right) }$ are the Lovelock tensors \cite{11}. After an
integration of the $tt$ component of the field equations (6) \cite{7} the
metric function is given by a third order ordinary equation 
\begin{equation}
24\alpha _{3}\left( 1-f\right) ^{3}+12\alpha _{2}\left( 1-f\right)
^{2}r^{2}+\left( 1-f\right) r^{4}-m=0
\end{equation}%
in which $m$ is the integration constant and relates to the mass of the
possible black hole solution. For our convenience let's introduce 
\begin{equation}
\frac{m}{r^{6}}=\mu ,\text{ \ \ }\frac{1-f}{r^{2}}=\mathcal{H},\text{ }%
12\alpha _{2}=\tilde{\alpha}_{2},\text{ \ }24\alpha _{3}=\tilde{\alpha}_{3}
\end{equation}%
which upon (8) the main equation (7) becomes 
\begin{equation}
\tilde{\alpha}_{3}\mathcal{H}^{3}+\tilde{\alpha}_{2}\mathcal{H}^{2}+\mathcal{%
H}-\mu =0.
\end{equation}%
The solution of this equation when both $\tilde{\alpha}_{2}$ and $\tilde{%
\alpha}_{3}$ are zero is just $\mathcal{H}=\mathcal{H}_{1}=\mu $ in which
the subscript $1$ denotes the first order Lovelock solution which is GR
solution. The GB solution is given if $\tilde{\alpha}_{3}=0$ but $\tilde{%
\alpha}_{2}\neq 0$ and therefore there are two solutions satisfying%
\begin{equation}
\tilde{\alpha}_{2}\mathcal{H}^{2}+\mathcal{H}-\mu =0
\end{equation}%
which are given by%
\begin{equation}
\mathcal{H}=\mathcal{H}_{2}^{\left( 1,2\right) }=\frac{-1\pm \sqrt{1+4\mu 
\tilde{\alpha}_{2}}}{2\tilde{\alpha}_{2}}.
\end{equation}%
Herein the subscript $2$ denotes the second order Lovelock solution or GB
solution. Imposing the condition that $\lim_{\tilde{\alpha}_{2}\rightarrow 0}%
\mathcal{H}_{2}=\mathcal{H}_{1}$ suggests that the only acceptable /
physical solution is the positive branch i.e., 
\begin{equation}
\mathcal{H}_{2}=\frac{-1+\sqrt{1+4\mu \tilde{\alpha}_{2}}}{2\tilde{\alpha}%
_{2}}.
\end{equation}%
We note that although $\tilde{\alpha}_{2}$ and $\tilde{\alpha}_{3}$ are two
real numbers but to avoid non-physical solutions an additional constraint
must be considered which is $1+4\mu \tilde{\alpha}_{2}\geq 0.$

Next, we consider $\tilde{\alpha}_{2}\neq 0$ and $\tilde{\alpha}_{3}\neq 0$
which in turn (7) admits three distinct solutions (one real and two complex)%
\begin{equation}
\mathcal{H}_{3}^{(1)}=\frac{1}{6\tilde{\alpha}_{3}}\left( \sqrt[3]{\Delta }+%
\frac{4\left( \tilde{\alpha}_{2}^{2}-3\tilde{\alpha}_{3}\right) }{\sqrt[3]{%
\Delta }}-2\tilde{\alpha}_{2}\right) ,
\end{equation}%
\begin{multline}
\mathcal{H}_{3}^{(2,3)}=-\frac{1}{12\tilde{\alpha}_{3}}\left( \sqrt[3]{%
\Delta }+\frac{4\left( \tilde{\alpha}_{2}^{2}-3\tilde{\alpha}_{3}\right) }{%
\sqrt[3]{\Delta }}+4\tilde{\alpha}_{2}\right) \pm \\
\frac{i\sqrt{3}}{12\tilde{\alpha}_{3}}\left( \sqrt[3]{\Delta }-\frac{4\left( 
\tilde{\alpha}_{2}^{2}-3\tilde{\alpha}_{3}\right) }{\sqrt[3]{\Delta }}\right)
\end{multline}%
in which%
\begin{equation}
\Delta =\eta +12\tilde{\alpha}_{3}\sqrt{3\xi },
\end{equation}%
where%
\begin{equation}
\eta =36\tilde{\alpha}_{2}\tilde{\alpha}_{3}+108\mu \tilde{\alpha}_{3}^{2}-8%
\tilde{\alpha}_{2}^{3}
\end{equation}%
and%
\begin{equation}
\xi =4\tilde{\alpha}_{3}-\tilde{\alpha}_{2}^{2}+18\tilde{\alpha}_{2}\tilde{%
\alpha}_{3}\mu +27\mu ^{2}\tilde{\alpha}_{3}^{2}-4\mu \tilde{\alpha}_{2}^{3}.
\end{equation}%
Here also the subscript $3$ denotes the third order Lovelock solution. Now,
similar to the GB solution, we require that the solutions (13)-(14) to give
the GB limit when $\tilde{\alpha}_{3}\rightarrow 0$. To do so we expand all
the three solutions (13)-(14) around $\tilde{\alpha}_{3}=0$ to find%
\begin{equation}
\mathcal{H}_{3}^{(1)}=-\frac{\tilde{\alpha}_{2}}{\tilde{\alpha}_{3}}+\frac{1%
}{\tilde{\alpha}_{2}}+\mathcal{O}\left( \tilde{\alpha}_{3}\right) ,
\end{equation}%
\begin{equation}
\mathcal{H}_{3}^{(2,3)}=\frac{-1\pm \sqrt{1+4\tilde{\alpha}_{2}\mu }}{2%
\tilde{\alpha}_{2}}+\mathcal{O}\left( \tilde{\alpha}_{3}\right) .
\end{equation}%
It is observed that $\mathcal{H}_{3}^{(1)}$ is singular at $\tilde{\alpha}%
_{3}=0$ while $\lim_{\tilde{\alpha}_{3}\rightarrow 0}\mathcal{H}_{3}^{(2,3)}=%
\mathcal{H}_{2}^{\left( 1,2\right) }$ which are the desired GB limits. It is
clear now why the known black hole solution in third order Lovelock gravity
does not give the GB limit. As a result, non of the solutions of the form
given in (13) and (14) can represent an acceptable solution in third order
Lovelock gravity with the correct GB limit. Perhaps the solution $\mathcal{H}%
_{3}^{(1)}$ is more physical than the other complex solutions and that is
why in many references, (13) was considered as the third order Lovelock
black hole solution in spherically symmetric spacetime.

Although the adopted solution in third order Lovelock gravity is $\mathcal{H}%
_{3}^{(1)}$ but one should not forget that it fails to give GB limit when $%
\tilde{\alpha}_{3}\rightarrow 0$. The same argument is also valid for the GB
black hole solutions $\mathcal{H}_{2}^{\left( 1,2\right) }.$ As we have
shown above, the positive branch of (11) yields the GR limit when $\tilde{%
\alpha}_{2}\rightarrow 0$ and therefore it is called physical solution while
the negative branch which has no GR limit is called exotic solution.

From the nature of the field equations, one has no hope to avoid such
ambiguity but here we try to find a solution which for small value of $%
\tilde{\alpha}_{3}$ satisfies the field equation (7) and correctly has the
GB limit. Let us reconsider the general equation (7) and introduce 
\begin{equation}
\mathcal{H}_{3}:=\mathcal{H}_{3}^{\left( 0\right) }+\tilde{\alpha}_{3}%
\mathcal{H}_{3}^{\left( 1\right) }+...
\end{equation}%
in which for small $\tilde{\alpha}_{3}$ we keep up to the linear term i.e.%
\begin{equation}
\mathcal{H}_{3}\simeq \mathcal{H}_{3}^{\left( 0\right) }+\tilde{\alpha}_{3}%
\mathcal{H}_{3}^{\left( 1\right) }.
\end{equation}%
A substitution in Eq. (7) and eliminating the higher order of $\tilde{\alpha}%
_{3}$ terms one finds 
\begin{multline}
\tilde{\alpha}_{3}\left( \mathcal{H}_{3}^{\left( 0\right) 3}+2\tilde{\alpha}%
_{2}\mathcal{H}_{3}^{\left( 0\right) }\mathcal{H}_{3}^{\left( 1\right) }+%
\mathcal{H}_{3}^{\left( 1\right) }\right) + \\
\tilde{\alpha}_{2}\mathcal{H}_{3}^{\left( 0\right) 2}+\mathcal{H}%
_{3}^{\left( 0\right) }-\mu \simeq 0.
\end{multline}%
A solution to this equation is found if%
\begin{equation}
\tilde{\alpha}_{2}\mathcal{H}_{3}^{\left( 0\right) 2}+\mathcal{H}%
_{3}^{\left( 0\right) }-\mu =0
\end{equation}%
and%
\begin{equation}
\mathcal{H}_{3}^{\left( 0\right) 3}+2\tilde{\alpha}_{2}\mathcal{H}%
_{3}^{\left( 0\right) }\mathcal{H}_{3}^{\left( 1\right) }+\mathcal{H}%
_{3}^{\left( 1\right) }=0.
\end{equation}%
The zero order condition (23) is the GB equation (10) and therefore $%
\mathcal{H}_{3}^{\left( 0\right) }=\mathcal{H}_{2}=\frac{-1+\sqrt{1+4\mu 
\tilde{\alpha}_{2}}}{2\tilde{\alpha}_{2}}.$ The first order condition (24)
also gives%
\begin{equation}
\mathcal{H}_{3}^{\left( 1\right) }=-\frac{\mathcal{H}_{3}^{\left( 0\right) 3}%
}{1+2\tilde{\alpha}_{2}\mathcal{H}_{3}^{\left( 0\right) }}
\end{equation}%
or consequently%
\begin{equation}
\mathcal{H}_{3}^{\left( 1\right) }=-\frac{\mathcal{H}_{2}^{3}}{1+2\tilde{%
\alpha}_{2}\mathcal{H}_{2}}.
\end{equation}%
Combining the results up to the first order of the third order Lovelock
parameter, the solution becomes%
\begin{equation}
\mathcal{H}_{3}\simeq \mathcal{H}_{2}\left( 1-\frac{\tilde{\alpha}_{3}%
\mathcal{H}_{2}^{2}}{1+2\tilde{\alpha}_{2}\mathcal{H}_{2}}\right) .
\end{equation}%
This technique, in principle, can be used for higher order of $\tilde{\alpha}%
_{3}$ but we shall not go through this here. The above solution has the
correct limit of GB when $\tilde{\alpha}_{3}=0$ and also for the case at
which $\tilde{\alpha}_{2}=0$ the solution is still valid but $\mathcal{H}%
_{2}=\mu .$

\section{Generalization to $d-$dimensional case with cosmological constant
and matter field}

The $d-$dimensional third order Lovelock theory of gravity coupled to a
matter source with cosmological constant is described by the following action%
\begin{multline}
I=\int d^{d}x\sqrt{-g}\times \\
\left( R-\frac{\left( d-1\right) \left( d-2\right) }{3}\Lambda +\alpha _{2}%
\mathcal{L}_{2}+\alpha _{3}\mathcal{L}_{3}+\mathcal{L}_{matt}\right) .
\end{multline}%
The static spherically symmetric line element is given by 
\begin{equation}
ds^{2}=-f\left( r\right) dt^{2}+\frac{dr^{2}}{f\left( r\right) }%
+r^{2}d\Omega _{d-2}^{2}
\end{equation}%
and $f\left( r\right) $ is given by the Einstein equations%
\begin{equation}
\mathcal{G}_{\mu \nu }^{\left( 1\right) }+\alpha _{2}\mathcal{G}_{\mu \nu
}^{\left( 2\right) }+\alpha _{3}\mathcal{G}_{\mu \nu }^{\left( 3\right) }-%
\frac{\left( d-1\right) \left( d-2\right) }{6}\Lambda g_{\mu \nu }=T_{\mu
\nu }
\end{equation}%
in which $T_{\mu \nu }$ is the energy-momentum tensor representing the
matter fields. Using $\frac{1-f}{r^{2}}=\mathcal{H}$, the $tt$ component of
the field equations (30) yields%
\begin{multline}
\tilde{\alpha}_{3}\mathcal{H}^{3}+\tilde{\alpha}_{2}\mathcal{H}^{2}+\mathcal{%
H}-\frac{\Lambda }{3}= \\
\frac{4m}{\left( d-2\right) r^{d-1}}-\frac{2}{\left( d-2\right) r^{d-1}}%
\dint r^{d-2}T_{t}^{t}dr
\end{multline}%
in which $T_{t}^{t}$ is the $tt$ component of the energy momentum tensor and 
\begin{equation}
\tilde{\alpha}_{2}=\left( d-3\right) \left( d-4\right) \alpha _{2}
\end{equation}%
\begin{equation}
\tilde{\alpha}_{3}=\left( d-3\right) \left( d-4\right) \left( d-5\right)
\left( d-6\right) \alpha _{2}.
\end{equation}%
Setting 
\begin{equation}
\mu =\frac{\Lambda }{3}+\frac{m}{r^{d-1}}-\frac{2}{\left( d-2\right) r^{d-1}}%
\int r^{d-2}T_{t}^{t}dr
\end{equation}%
in Eq. (31), we refined the previously discussed equation (9). This means
that the $d-$dimensional case with cosmological constant and matter field
reduces to the case of $7-$dimensional, providing $\mu $ is given by (34).

\begin{figure}[tbp]
\includegraphics[width=60mm,scale=0.7]{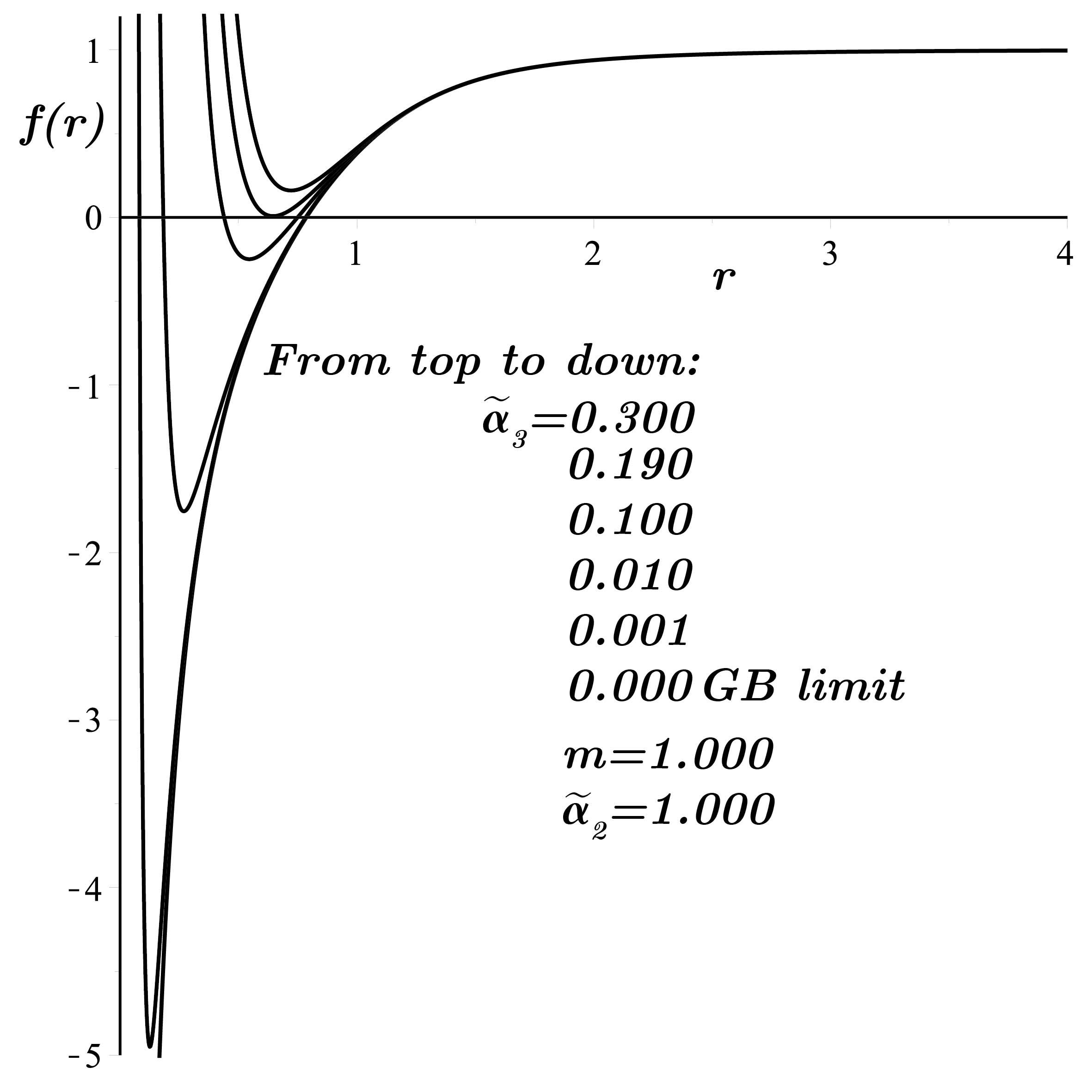}
\caption{Plots of $7-$dimensional metric function $f\left( r\right) =1-r^{2}%
\mathcal{H}_{3}$ in terms of $r$ for $m=1,$ $\tilde{\protect\alpha}_{2}=1$
and various values of $\tilde{\protect\alpha}_{3}.$ The form of $\mathcal{H}%
_{3}$ is approximated in Eq. (27).}
\end{figure}

\section{Discussion}

The third order Lovelock theory of gravity admits static spherical symmetric
black hole solution in $d\geq 7$ dimensions which unexpectedly does not
admit the GB black hole solution when the third order Lovelock parameter
vanishes. An approximation solution which is valid for small $\tilde{\alpha}%
_{3}$ has been found. This new solution gives the correct GB limit and up to
the first order of $\tilde{\alpha}_{3}$ satisfies the field equations. In
Fig. 1 we plot the metric function $f\left( r\right) $ of the third order
Lovelock gravity for the different values of $\tilde{\alpha}_{3}$ in $7-$%
dimensions. As depicted in Fig.1 the effect of $\tilde{\alpha}_{3}\neq 0,$
even if it is small, is not trivial and the structure of the spacetime
changes.

\end{document}